\documentclass[aps,twocolumn,amsfonts]{revtex4}
\begin{document}
\title{Quantum Theory of Tensionless Noncommutative
$p$-Branes}
\author{J.  Gamboa$^{1}$\thanks{E-mail:jgamboa@lauca.usach.cl},
M.~Loewe$^{2}$ \thanks{E-mail:mloewe@fis.puc.cl} and
F.  M\'endez$^{3}$\thanks{E-mail:fernando.mendez@lngs.infn.it}}
\affiliation{$^{1}$Departamento   de   F\'{\i}sica,   Universidad   de
Santiago de Chile, Casilla 307, Santiago 2, Chile \\ $^{2}$Facultad de
F\'{\i}sica, Pontificia Universidad  Cat\'olica de Chile, Casilla 306,
Santiago  22,  Chile \\  $^{3}$INFN,  Laboratorio  Nazionali del  Gran
Sasso, SS, 17bis, 67010 Asergi (L'Aquila), Italy}

\begin{abstract}
The  quantum  theory   involving  noncommutative
tensionless  $p$-branes is  studied following  path  integral methods.
Our  procedure  allow  a  simple  treatment  for  generally  covariant
noncommutative extended systems and it contains, as a particular case,
the  thermodynamics and  the  quantum tensionless  string theory.  The
effect induced by noncommutativity in  the field space is to produce a
confinement among pairing of null $p$-branes.
\end{abstract}
\maketitle
%\pacs{ PACS numbers:03.65.-w, 03.65.Db}

\section{Introduction}

Tensionless strings are extended  objects discovered by Schild many
years ago  \cite{sch} and  it corresponds --formally-  to a  set of
infinite massless relativistic  particles satisfying the constraint

\[ {\cal H}_1 = p_\mu x^{'\mu}, \]

where  $x^\mu$ transform  as  a scalar  on  the volume  world and  the
spacetime index $\mu$ runs over $0, 1, 2, ...,D-1$ with $D$,  the
spacetime dimension\footnote{For recent results results see \cite{varios}}.

Physically  speaking this problem  is related  to the  behavior of
string theory at very high energy \cite{witte}, also known as
the strong coupling limit.  More exactly, when the Regge slope goes
to  infinite, the  spectrum  of  the string  theory  is massless;  the
situation is similar  to what occurs in the  standard model before the
gauge symmetry is broken.

In string theory, however, the situation is quite involved because
the \lq \lq gauge group" corresponds to diffeomorphisms and the
gauge group --in  this case is infinite  dimensional. As in the
standard string theory, the critical dimensions for space time
will be 26 (or 10) depending on the bosonic (or
fermionic)character \cite{gamboa} as in the tensionful case.

The  general case  (that is  $p$-branes), concerning  to  the critical
dimensions, only  partial results are  known and, probably,  they are
not definitive \cite{scho}.

The fact that null strings exist  at very high energies, in the sense
previously explained,  together with the possibility  that the Lorentz
invariance  could  be  deformed  or  even  broken  at  such  energies
\cite{kos,alot,ame1,ame2} rise the question of how to study the effect
of such deformation in the null string scenario.

There are several proposals to formulate such Lorentz
invariance deformations.
They can be classified in two groups depending on the existence of
a preferred reference frame. For such a case there is a  proposal
\cite{ccgm,rubi}  based   in  the  deformation   of  the
commutators between fields which is appropriated to discuss the
question previously formulated.

However, one  could argue  that such deformation  should not  have any
impact  in the sense of  a measurable consequence.  That is, the
deformation at  the level  of just one  string or $p$-brane,  could be
washed out. Then, there should be an amplifier mechanism.

Because of this, instead of considering just one $p$-brane, turns out
to be more interesting to consider  a gas of such objects and to study
the  thermodynamics  of  such  system.  Then, one  can  introduce  the
noncommutative fields and explore  the amplified consequences of
it.

The purpose of the present paper is to study the properties of these extended objected described by noncommutative fields and as well as some statistical mechanics of  null $p$-branes issues that 
 include  the null string is as a particular case.

The  paper  is organized  as  follow:  in  section II  we  consider
relativistic  particles  in a  noncommutative  space where  several
quantum  statistical  mechanics   considerations  are  studied.  In
section  III,  the previous  results  are extended  non-commutative
$p$-dimensional null  branes and its  quantum statistical mechanics
properties described. In section IV, we discuss the previous results
as a  possible interaction mechanism  and also we present  our main
conclusions. An appendix including the statistical mechanics for free relativistic particles is considered and the nonexistence of Matsubara modes for this system is established. 

\section{Noncommutative relativistic quantum mechanics}
\label{sec:NC1}

In  this  section  we  will construct  noncommutative  versions  of
generally  covariant systems. We  will start  considering, firstly,
the  relativistic  particle  on  a  $D$-dimensional  spacetime  and
later--  in the  next  section --  we  will extend  our results  to
tensionless strings and membranes.

\subsection{Relativistic free particle and the proper-time gauge}

There are many approaches to discuss relativistic quantum mechanics
of  a free  particle.  One of  them  is the  so called  proper-time
method, which was used in the early 50th in connection with quantum
electrodynamics \cite{nambu}. The idea is to consider a particle in
a $D+1$- dimensional Euclidean spacetime.

The diffusion equation for such system is
\begin{equation}
-\frac{1}{2} \Box \varphi (x, s) = \frac{\partial \varphi}
{\partial s},
\label{6}
\end{equation}
where  $\Box$ is  the $D$-dimensional   Laplacian.

Then, using the ansatz
\begin{equation}
\varphi (x,s) = e^{-\frac{m^2}{2} s} \phi (x),
\label{7}
\end{equation}
one finds that $\phi(x)$ satisfies the Klein-Gordon equation if $m$ is
the mass of the particle.

In this  approach, the propagation  amplitude is given by  the Laplace
transform
\begin{equation}    G[x,x^{'};    m^2]    =    \int_0^\infty    ds
e^{-s\frac{m^2}{2} } \, G[x,x^{'}; s],
\label{8}
\end{equation}
where
\begin{eqnarray}
G[x,x^{'}; s] &=&  \int {\cal D} x \,  e^{-\int_0^1 d \tau \frac{{\dot
x}^2}{2 s}}, \label{9} \nonumber
\\
&=&s^{-D/2} \, e^{- \frac{(\Delta x)^2}{2 s}}.
\label{99}
\end{eqnarray}

From this one obtains the partition  function for a gas of $N$ free
relativistic particles  \footnote{Along  this paper we  assume that the particles are spinless. The  reader  should note  also  that we  are assuming the Maxwell-Boltzmann statistics, for a justication about this see appendix}
\begin{equation}
Z_s=\left(\mbox{Tr}\left[e^{-\frac{m^2}{2}  s}
G[x,x';s]\right]\right)^N,
\end{equation}
or    equivalently
\begin{equation}
\ln Z = N  \left[ - \frac{m^2}{2} s- \frac{D}{2} \ln s  + \ln {\cal V}
\right],
\label{10}
\end{equation}
where ${\cal V}=V \times \mbox{const.}$ is the $D$-dimensional spacetime,
$V$ is the $D-1$-dimensional  ordinary spatial volume and $s$ plays
the role of $\beta =1/kT$.

\subsection{The relativistic particle in a noncommutative space}

Equation (\ref{6}) suggests a simple way to extent the problem to a
gas of relativistic particles on a noncommutative space.

Indeed,  from   (\ref{6})  we  see  that  the   Hamiltonian  for  a
relativistic particle is
\begin{equation}
{\hat H} =\frac{1}{2} p^2_\mu.
\label{ham}
\end{equation}

Once (\ref{ham}) is  given, noncommutativity is implemented through
the deformed algebra
\begin{eqnarray}
\left[     x_\mu,      x_\nu\right]&=&     i     \theta_{\mu     \nu},
\,\,\,\,\,\,\,\,\,\,\,   \left[  p_\mu,p_\nu\right]  =i   B_{\mu  \nu},
\label{11}
\\
\left[ x_\mu,p_\nu\right] &=& i \delta_{\mu \nu},
\label{12}
\end{eqnarray}
where  $\theta_{\mu \nu}$  and  $B_{\mu \nu}$  are the  deformation
parameters in the phase space.

For convenience we choose
\begin{eqnarray}  \theta_{i0}   &=&  0\,  \,\,\,\,\,\,\,\theta_{ij}  =
\epsilon_{ij} \theta, \label{13}
\\
B_{i0} &=& 0, \,\,\,\,\,\,\,\,B_{ij} =\epsilon_{ij} B.
\label{14}
\end{eqnarray}

Therefore,   the  equation   of   motion  for   this  particle   is
\begin{eqnarray}
{\dot x}_\mu &=& p_\mu, \nonumber
\\
{\dot p}_i &=& \epsilon_{ij} B p_j.  \label{15}
\end{eqnarray}

These equations can be  integrated directly by using (\ref{13}) and
(\ref{14}). Indeed,  one of the  equations is trivial,  namely, the
energy  conservation  condition   ($\dot{p}_0=0$).  Note  that  the
symmetric gauge  we have chosen, implies  that non-commutativity is
realized   only  for   the  first   two  momenta   and  coordinates
components.  The   other  components  are  treated   as  usual.  In
principle, we  could extend this hypothesis taking  also other pairs
of momenta  and coordinates components,  but this is  not essential
for our discussion.

Keeping  this  in mind,  the  remaining  equations  have the  solution
\begin{eqnarray}
p_1 &=& \frac{1}{2}  \left( \alpha~ e^{-i B t} +\alpha^\dag  ~ e^{ i B
t} \right), \nonumber
\\
p_2 &=& \frac{1}{2i}\left(  \alpha ~e^{-  i B  t} -  \alpha^\dag~ e^{  i B  t} \right),
\label{16}
\end{eqnarray}
where $\alpha$'s are constant operators.

The  coordinates $x_  {1,2}$ are  obtained in  a similar  way using
(\ref{15}), {\it i.e.}
\begin{eqnarray}
x_1 &=& \frac{1}{2 i B}\left(\alpha^\dag~ e^{ i B t} -\alpha~ e^{- i B
t} \right) + x_{01}, \nonumber
\\
 x_2 &=& \frac{1}{2 B} \left( \alpha~  e^{- i Bt} + \alpha^\dag~ e^{ i
B t} \right)+ x_{02}.
\label{17}
\end{eqnarray}

From the commutation relation of  $p$'s, we see that it is possible
to define  operators $a$  and $a^{\dagger}$ satisfying  the algebra
\begin{eqnarray}
\left[a, a \right] &=& 0=\left[a^\dag, a^\dag\right], \nonumber
\\
\left[a, a^\dag \right] &=& 1 \label{18},
\end{eqnarray}
where
\[ \alpha \rightarrow \sqrt{B} a,\,\,\,\,\,\,\,\,\,\,\,\,\,\,\,\,\,\, \alpha^{\dag} \rightarrow
\sqrt{B}a^{\dag}.  \]

The equations of  motion --as a second order  equation system-- are
\[ \ddot{x}_\mu=B_{\mu\nu}\dot{x}_\nu, \]
which can be solved by the Ansatz $x_\mu = a_\mu \,e^{i\omega s}$.

The last equation is
\[ (i\omega\delta_{\mu\nu}-B_{\mu\nu})a_\nu=0.  \]

Therefore,   the   dispersion   relation   for   this   system   is
\begin{equation}
\omega_{\pm}=\left\{
\begin{array}{c} \pm B  \\ 0
\end{array} \right. , \label{dis}
\end{equation}
and --since one of the eigenvalues vanishes-- the Hamiltonian spectrum is degenerated

Thus,  the hamiltonian  for  a relativistic  particle  living on  a
noncommutative space is
\begin{equation} H=\frac{B}{2} \left( a^\dag a +\frac{1}{2}\right) + \frac{1}2\sum_{n=1}^{D-3}(p_\mu^2)_n.
\label{23}
\end{equation}

Finally, the  statistical mechanics for  a gas of  $N$ relativistic
particles  on a noncommutative  space, in  the symmetric  gauge, is
obtained from the partition function
\begin{eqnarray}
Z_s&=&\left(s^{-\frac{D-3}{2}}e^{-\frac{m^2}{2}s} \sum_{n=0}^{\infty}
{\cal G}_0 e^{-s\frac{B}{2}(n+\frac{ 1}{2})}\right)^N,\nonumber
\\
&=&\left[\frac{{\cal              G}_0             e^{-\frac{m^2}{2}s}
  s^{-\frac{D-3}{2}}}{\sinh{(\frac{B}{2} s)}} \right]^N,
\label{25}
\end{eqnarray}
where  ${\cal G}_0$ is the degeneracy  factor due to
the  zero  eigenvalue of  the  Hamiltonian \footnote{Although  this
factor can be computed by using a regularization prescription, here
this factor is absorbed as a normalization constant.}.

The  thermodynamic  properties  of  this  system  can  be  computed
directly from (\ref{25}).

\section{The strong coupling regime for membranes in noncommutative
 spaces}
\label{sec:NC2}

In  this section  we will  discuss  the extension  of the  previous
problem to membranes moving on a noncommutative space in the strong
coupling regime.

A relativistic  membrane is a $p$-dimensional object  embedded on a
$D$-dimensional  flat  spacetime and  described  by the  lagrangean
 density
\[  {\cal  L} =  \frac{T}{2}  \sqrt{g^{(p+1)}} \left[g_{\alpha  \beta}
G^{\mu  \nu}  \partial^{\alpha}  x_\mu \partial^{\beta}  x_\nu  -(p-1)
\right], \]
where $g^{(p+1)}_{\alpha  \beta}$ ($\alpha, \beta= 0, 1,  2, ...p$) is
metric tensor on the world-volume,  a$G^{\mu \nu}$  is the  metric tensor where  the $p$-brane  is embedded with $\mu, \nu = 0,1,2, ...,D$ and $T$ the superficial tension.

The  hamiltonian  analysis  yields  to  the  following  constraints
\begin{eqnarray}
H_{\perp}&=&\frac{1}{2}(p^2 + T^2 g^{(p)}),
\label{26}
\\
H_i&=&p_\mu\partial _i x^\mu, \label{27}
\end{eqnarray}
where  $g^{(p)}$ is  the spatial  metric  determinant and  $T$ is  the
superficial tension.

The strong coupling regime  corresponds to $T\rightarrow 0$ and, in
this limit the constraints are
\begin{eqnarray}
H_{\perp}&=&\frac{1}{2} p^2 ,\label{2666}
\\ H_i&=&p_\mu\partial _i x^\mu, \label{2777}
\end{eqnarray}
 and  the  membrane becomes  an
infinite   set  of   free  massless   relativistic   particles  moving
perpendicularly to the $p$-dimensional surface.

In the special  case of the tensionless string  ($p=1$), each point
of the  string is associated with a  massless relativistic particle
and, as  a consequence, all the  points of the  string are causally
disconnected.

In this tensionless string approach the field $x^\mu(\sigma ,\tau)$
is  replaced  by $x^\mu_i  (\tau)$,  where  $i=1,  2, ...,$  is  an
infinite  countable  set labeling  each  point  of the  tensionless
string.

Using  this  philosophy,  we  will start  constructing  tensionless
 strings.

\subsection{Tensionless strings from particles}

Let us start by noticing that a tensionless string \cite{gamboa} is
made  up  of  infinite  massless  relativistic  particles  causally
disconnected  and,   therefore,  instead  of   (\ref{6})  one  have
\begin{eqnarray}
-\frac{1}{2}\Box     \varphi_1      (x,     s_1)&=&     \frac{\partial
\varphi_1}{\partial s_1}, \nonumber
\\
-\frac{1}{2}\Box\varphi_2   (x, s_2) &=&
\frac{\partial\varphi_2}{\partial s_2},\nonumber \\
 &\vdots& \nonumber
\\
-\frac{1}{2}\Box\varphi_k (x, s_k)
&=&\frac{\partial\varphi_k}{\partial s_k}.
\end{eqnarray}

These equations can be solved by generalizing the Ansatz (\ref{7}),
{\it i.e}
\begin{equation}
\varphi(x_1,\dots,x_k,\dots;s_1\dots,s_k,\dots)=\prod_{i=1}^{\infty}~e^{-\frac{m^2}{2}s_i}\phi(x_i),
\end{equation}
where  $m^2$ is  an  infrared  regulator that  will
vanish at the end of the calculation.

The limit  of an  infinite number of  particles is delicate  but here
--formally-- one can take this limit, simply, assuming that in the
continuous limit one can replace the  set $\{ i\}$ by an integral in
$\sigma$  and, as a  consequence, the  propagation amplitude  can be
written as:
\begin{eqnarray}
&G&[x(\sigma),x'(\sigma)]=\nonumber
\\
&=&  \int_0^{\infty}  {\cal D}s(\sigma)~e^{-\frac{m^2}{2}\int  d\sigma
s(\sigma)}G[x(\sigma), x'(\sigma);s(\sigma)], \label{29}
\end{eqnarray}
where $G[x(\sigma),x'(\sigma);s(\sigma)]$ is given by
\begin{equation}
G[x(\sigma),x'(\sigma);s(\sigma)]=s^{-D/2}(\sigma) e^{-\int d\sigma
\frac{[\Delta        x(\sigma)]^2}{2s(\sigma)}}.        \label{299}
\end{equation}

The  formula  (\ref{29}) generalizes  the  proper-time  method to  the
tensionless string  case. Probably this approach to  string theory was
first used by Eguchi in \cite{gamboa}.

Using (\ref{29}) and (\ref{299}),  the partition function of an $N$
tensionless string gas is
\begin{eqnarray}
Z[s(\sigma)]&=&\left[\int                   {\cal                   D}
      x(\sigma)\,G[x(\sigma),x(\sigma);s(\sigma)]\right]^N\nonumber
\\
&=&\left(s^{-D/2}~e^{-\int  d\sigma  \frac{m^2}{2}s(\sigma)}\right)^N.
\label{32}
\end{eqnarray}

This partition  function reproduces  correctly the results  for the
thermodynamics of a tensionless string gas \cite{string}.

Indeed, from (\ref{32}), the Helmholtz free energy is
\[ F[s]=\frac{N}{s(\sigma)}\left[ \frac{D}{2} \ln(s(\sigma))
+\frac{m^2}{2}\int d\sigma s(\sigma) +\ln({\cal V}) \right].  \]

As $1/s$ is the temperature,  then from the limit $m^2\rightarrow 0
$ we see that $F/T \sim \ln(T)$, again in agreement with other null
string calculations \cite{string,atick}.

From the last equation one obtain that
\begin{equation}
P[s(\sigma)]{ V}=\frac{N}{s(\sigma)},
\end{equation}
is the state equation for an ideal tensionless string gas.

\subsection{Tensionless   membranes   from   tensionless   strings}
\label{sec:NC3}

In order to construct tensionless membranes, we begin by considering
a membrane as an  infinite collection of tensionless strings.  Thus,
if the  membrane is a  p-dimensional object, with  local coordinates
$(\sigma_1,\dots,\sigma_p)$,   then   the   propagation   amplitude,
formally, corresponds to (\ref{29}), with the substitution

\[ \sigma\rightarrow (\sigma_1,\dots,\sigma_p).  \]

Therefore,  the partition  function for  a gas  of  $N$ tensionless
membranes is
\begin{eqnarray}
Z[s(\sigma)]= \left[\lim_{n \to\infty}\left([s(\sigma)]^{-D/2}~
e^{-     \frac{m^2}{2}\int    d^p     \sigma     s(\sigma)}{\cal    V}
\right)^n\right]^N, \nonumber \label{34}
\\
&&
\end{eqnarray}
where $n$ is the number of tensionless strings.

One should note here that the expression

\[\left([s(\sigma)]^{-D/2}~e^{-    \frac{m^2}{2}\int    d^p    \sigma
   s(\sigma)}\right)^n, \]

formally  emphasizes that  a  tensionless
$p$-branes is made-up of $n$ tensionless strings.

However, this  last expression was  computed in (\ref{299})  and in
our case is
\[  \prod_{i=1}^p[s(\sigma_i)]^{-D/2}   \,  e^{-\frac{m_i^2}{2}\int  d
\sigma_i s(\sigma_i)}, \]
then, the total partition function for an ideal gas of $N$ tensionless
p-branes is given by
\[ Z = \prod_{i=1}^p \left([s(\sigma_i)]^{-D/2}~ e^{-\frac{1}{2}m_i^2
\int d\sigma_i s(\sigma_i)}\right)^N. \]

In order to  compute the state equation we  proceed as follow: firstly
one chooses  $s(\sigma_1)=s(\sigma_2) ...=s(\sigma)$ and  one put also
$m_1=m_2= ...=m$, then
\begin{equation}
P[s(\sigma)]~{ V} =\frac{N}{s(\sigma)}.
\end{equation}

The Helmholtz free energy, compared to the tensionless string case,
has a different behavior. Indeed, the Helmholtz free energy becomes

\[  F[s]=\frac{pN}{s(\sigma)}\left[  \frac{D}{2}  \ln(p\,s(\sigma))
+\frac{m^2}{2}\int d\sigma s(\sigma)  +\ln({\cal V}).  \right].  \]
and for $s\rightarrow  \infty$, one has that the  quantity $sF \sim
\frac{D}{2} \ln [p\, s]$ is similar to the string case but, in this
case $p$ could smooth out the behaviour of $sF$.

\subsection{Including noncommutativity in Tensionless p-branes}

Using the previous  results, we can generalize our  arguments in order
to include  noncommutativity in tensionless  p-branes. In order  to do
that,  one start considering  a tensionless  p-brane described  by the
{\bf  field} $x_i^\mu  (\tau)$  with $i$  labeling  the dependence  in
$(\sigma_1,\sigma_2, ..\sigma_p)$.  This  field transforms as a scalar
on the world-volume but as a  vector in the space where the p-brane is
embedded.

Let  us  suppose  that  the  components  --we  say  $x_i^{{D-1}}$  and
$x_i^D$-- do not  commute, then -in such case-  the Green function can
be written as
\begin{eqnarray}
&& G[x(\sigma),x'(\sigma);s(\sigma)] = \nonumber
\\
&=& \int_0^\infty ds\, e^{-\frac{m^2}{2} s}\prod_{k=0}^{D-3} \left[ \int {\cal D} x_i^k
\,  e^{-\int_0^1d   \tau  \,\frac{1}{2s}  ({\dot   x_i^k})^2  }\right]
\nonumber
\\
&\times&   \int  {\cal   D}  {x_i}^{(D-2)}   {\cal  D}
{x_i}^{(D-1)}  \,   e^{-\int_0^1  d  \tau   \frac{1}{2s}  \left(({\dot
x_i^{(D-2)}})^2 + ({\dot x_i^{(D-1)}})^2 \right)}. \nonumber \\
\label{gr1}
\end{eqnarray}

The integral in the second line  in the RHS, corresponds formally to a
non-relativistic particle with mass  ($s^{-1}$) moving in plane in the
presence of a constant perpendicular magnetic field $B$. In the first
line  in the  RHS, however,  the integral  formally correspond  to the
Green function for a set  of $p$ free relativistic particles moving in
$(D-3)$-dimensional spacetime.

Thus, the calculation of these integral is straightforward. Indeed,
\begin{eqnarray}
 G[x(\sigma),x'(\sigma)] &=& \nonumber
\\
= \int_0^\infty    ds   [s(\sigma)]^{-\frac{D-3}{2}}&~&s(\sigma)   \,e^{
-\frac{1}{2s}(\Delta x^k_i)^2 - \frac{p}{2}m^2 \int d\sigma s(\sigma)}
\nonumber \\
&\times& \mbox{H. O.}, \nonumber
\end{eqnarray}
where $\mbox{H. O}$ means  the harmonic oscillator calculation for the
two-dimensional relativistic Landau problem.

The partition  function  for this  gas  of $N$-tensionless  $p$
branes
\begin{eqnarray}
&Z&[s(\sigma)]=\mbox{Tr}\left[G[x(\sigma),x'(\sigma);s(\sigma)]
\right]\nonumber
\\
&=&\biggl([s(\sigma)]^{-\frac{D-3}{2}}e^{-\frac{pm^2}{2}\int   d\sigma
s(\sigma)     }     \sum     _{n=0}^{\infty}{\cal     G}_0     e^{-
\frac{B}2(n+\frac{1}{2}\int   d^p\sigma   s(\sigma))}   \biggr)^{N}
\nonumber
\\
&=&\left[\frac{{\cal G}_0[s(\sigma)]^{-\frac{D-3}{2}}}{\sinh\left
(\frac{p\,B}{2}\int d\sigma s (\sigma)\right)}\right]^{N}.
\label{la}
\end{eqnarray}

Therefore, if we  assume pairing interaction, then noncommutativity
induces  a  motion for  a  tensionless  $p$-branes  confined via  a
harmonic potential oscillator.

\section{Interactions  via  noncommutativity  in the  phase  space}
\label{sec:NC4}

In the  previous section we argued how  to construct noncommutative
extended objects.  In this section we would like to give an insight
in a different physical  context and to investigate the possibility
of   a  possible  interaction   by  means   noncommutativity.  This
procedure, is simple extension of the noncommutative field.

From  the non-relativistic point  of view,  apparently there  is no
problem  with nonlocal communication  \cite{peres}. Indeed,  let us
suppose two non-relativistic particles in one dimension, labeled by
coordinates $x_1$  and $y_1$ and canonical momenta  $p_1$ and $p_2$
respectively.  Note that  the index  refers now  to  the particles
involved.

The Hamiltonian for this system is
\begin{equation}
H=\frac{1}{2}p_1^2 + \frac{1}{2}p_2^2.
\label{38}
\end{equation}

Although  naively  the particles  in  (\ref{38})  are  free, they  can
interact if we posit the commutator
\begin{equation}
[p_1,p_2]=iB,
\label{39}
\end{equation}
where $B$  measures the  strength of this  interaction which  can play
--or not--the role of a magnetic field.

The exact equivalence between this system and the Landau problem is
a subtle  point because by considering only  a noncommutative phase
space with noncommutative parameters $\theta$ and $B$, one can show
that  noncommutative  quantum  mechanics  and  the  Landau  problem
coincide if the  relation $\theta = 1/B$ is  fulfilled, i.e.  if we
have just  the magnetic  length \cite{gamboa1}. From  this example,
one extract  as conclusion that the equivalence  between a physical
system  such  as  the  Landau problem  and  noncommutative  quantum
mechanics only occurs for the critical point $\theta B =1$. For other values  of $\theta B$,  noncommutative quantum mechanics
describes a physics completely different from the Landau problem.

The above example can be generalized for more particles; for instance,
let us consider two free particles moving in a commutative plane.

The Hamiltonian is
\begin{equation}
H=\frac{1}{2}(p_{1x}^2+p_{1y}^2)+\frac{1}{2}(p_{2x}^2+p_{2y}^2).
\end{equation}

Then, let us  assume that the interaction is  given by \footnote{Of
course  this is a  simplification because  we are  assuming  that the
noncommutative parameters are the same.}
\begin{equation}
[p_{1x},p_{2x}]=iB,\,\,\,\,\,\,\,\,\,\,\,\,\,\,\,[p_{1y},p_{2y}]=iB,
\label{40}
\end{equation}
then, as in the previous case, the Hamiltonian is
\begin{equation}
H=\frac{1}{2}(p_{1x}^2+p_{2x}^2)+\frac{1}{2}(p_{1y}^2+ p_{2y}^2).
\label{41}
\end{equation}

Thus,  the  commutator (\ref{40})  and  the hamiltonian  (\ref{41})
describe  a couple  of particles  living  on a  plane and  interacting
formally with a magnetic field perpendicular to the plane.

We  would  like  to  remark  that our  procedure,  of  course,  has
generated a nonlocal interaction between both particles.

In the general  case for $N$ particles moving  on a $D$ dimensional
commutative space, the generalization is straightforward.

Indeed, the Hamiltonian is
\begin{equation}
H  =  \frac{1}{2}(p^2_{1x} +p^2_{1y}+ ...) + \frac{1}{2} (p^2_{2x} +p^2_{2y}+ ...) +
   ... ,
\label{l6}
\end{equation}
then the interaction can be written
as
\begin{equation}
[ p_i^a, p_j^b] = i \delta_{ij} \epsilon^{ab} B,
\label{l7}
\end{equation}
where $a,b$ run on $1, ...,N$ labeling the different species of particles and the indexes $i,j,
 ...$ select the vectorial component of ${\bf x}$ \footnote{The
components of  the antisymmetric density  tensor $\epsilon^{ab}$ are
defined as $+1$ if $a>b$.}.

If we rewrite the Hamiltonian as
\begin{equation}
H = \frac{1}{2}(p^2_{1x} +p^2_{1x}+ ...) + \frac{1}{2} (p^2_{1y} +p^2_{1y}+ ...) +
   ... .
\label{l8}
\end{equation}

Thus, in the critical point,  this generalized system is related to
the  quantum Hall  effect as  has been  proposed using  a different
argument by \cite{poly2}.

Thus,  in our  context, one  could conclude  that if  two particles
interact  via  nonlocal communication,  the  phase  space could  be
noncommutative. However, this fact  does not exclude other possible
mechanisms as a source of nonlocal interactions.

\section{Conclusions}

In  conclusion, we  have constructed  the statistical  mechanics of
generally covariant systems such as $p$-branes assuming that for each
point of  the world-volume one define a  noncommutative field. From
these results, we have studied the quantum statistical mechanics of
tensionless p-branes gas that  is a qualitatively different system in
comparison with the commutative one.

In  addition,  we  have  discussed  a  possible  mechanism  for  to
implement no-local  interactions by means  of noncommutativity that
could be useful in the quantum Hall effect or other systems.

The possible cosmological implications of these results as well as
other results are also studied in \cite{das}.

\acknowledgments   We   would  like   to   thank   J.  L.    Cort\'es,
A. P.  Polychronakos and J. C.  Retamal by discussions.  This work has
been partially  supported by the  grants 1010596, 1010976 from Fondecyt. F. Mendez thanks to INFN for a for postdoctoral fellowship. 

\appendix 

\section{statistical mechanics for a free relativistics particles gas}

In this appendix we will study the statistical mechanics for a free  relativistics particles gas. One start considering the partition function 
\begin{equation}
Z = Tr ~G\left[x_2,x_1\right], \label{palma1}
\end{equation}
where $G\left[x_2,x_1; m^2 \right]$  is the Green function for a free particle in the sense discussed in section II and it is given by
\begin{equation}
G [x_2, x_1; m^2] = s^{-\frac{D}{2}} e^{-\frac{(\Delta x)^2}{2 s}}, \label{app1}
\end{equation}
where $D$ is the dimension of the spacetime. 

In the euclidean space (\ref{palma1}) is computed using periodic boundary conditions but instead of one use 
\begin{eqnarray}
x^{0}_1 (0) &=& x^{(0)}_2 (T) + 2 n \pi R, \label{appe1a}
\\
x^{D-1}_1 &=& x^ {D-1}_2  (T).  \label{appe1b}
\end{eqnarray}
where $i=1, 2, ..., D-1$, $n=0,\pm1,\pm2,...$ are the Matsubara frequencies and $R$ is the compactification radius.

Using this fact, one find that the total partition function is 
\begin{eqnarray}
Z&=& \sum_{n=1}^\infty Z^{(n)}, \label{part1}
\\
&=& \sum_{n=1}^\infty ~e^{i\,n \theta} Z^{(n)},\label{par21}
\\
&=& s^{-D/2} \vartheta_3 \left( e^{-\frac{2\pi^2 R^2}{s}}\right)~e^{i\theta}, \label{par31}
\end{eqnarray}
where $\vartheta_3$ is the Jacobi function and $\theta$ is phase factor that plays the analogous role of the magnetic flux in the Aharonov-Bohm effect. Since $s$ play an analog role of $\beta$ in statistical mechanics, in the high temperatures limit  $\vartheta_3 \rightarrow 1$ and the logarithm of the partition function in this case is 
\begin{equation}
\ln Z = N  \left[ - \frac{m^2}{2} s- \frac{D}{2} \ln s  + \ln {\cal V} 
\right].
\label{1000}
\end{equation}

Thus, one find that in the spinless case and in the high temperature region, there are no Matsubara modes and the statistics is, of course, the Maxwell-Boltzmann one.

\end{document}